\documentclass[11pt]{article}
\usepackage{jheppub} 

\usepackage{braket,slashed,bm}
\usepackage{array,multirow}
\usepackage{simplewick}
\usepackage[T1]{fontenc} 

\usepackage{tikz}
\usetikzlibrary{arrows,decorations.pathmorphing,backgrounds,positioning,fit,petri,automata,shadows,calendar,mindmap,
decorations.markings,calc}

\def\nn{\nonumber\\ }
\def\rd{{\rm d}}

\def\lsix{ \mathcal{L}^{(6)}}

\def\msbar{$\overline{\hbox{MS}}$}
\def\hyp{\mathsf{y}}

\allowdisplaybreaks[0]

\title{
Renormalization Group Evolution of the Standard Model Dimension Six Operators\\
I: Formalism and  $\lambda$ Dependence
}

\author[a]{Elizabeth E.~Jenkins,}

\author[a]{Aneesh V.~Manohar,}

\author[b,1]{Michael Trott}\note{Corresponding author.}

\affiliation[a]{Department of Physics, University of California at San Diego, 9500 Gilman Drive,\\ La Jolla, CA 92093-0319, USA}
\affiliation[b]{Theory Division, Physics Department, CERN, CH-1211 Geneva 23, Switzerland}

\emailAdd{ejenkins@ucsd.edu}
\emailAdd{amanohar@ucsd.edu}
\emailAdd{michael.trott@cern.ch }

\abstract{
We calculate the order $\lambda$, $\lambda^2 $ and $\lambda y^2$ terms of the $59 \times 59$ one-loop anomalous dimension matrix of
dimension-six operators, where $\lambda$ and $y$ are the Standard Model Higgs self-coupling and a generic Yukawa coupling, respectively.
The dimension-six operators modify the running of the Standard Model parameters themselves, and we compute the complete one-loop result for this. We discuss how there is mixing between operators for which no direct one-particle-irreducible diagram exists, due to operator replacements by the equations of motion.
}

\begin{document}
\maketitle

\section{Introduction}\label{sec:intro}

The LHC has discovered a Higgs-like boson with a mass of approximately 126~GeV, with properties consistent with the standard model to within current experimental errors. The Standard Model (SM) also provides a good description of all the LHC data to date, with no evidence for beyond the SM (BSM) physics. The current experimental results can be described by the Standard Model with a scalar doublet which spontaneously breaks the gauge symmetry, and with BSM physics parameterized by higher dimension operators constructed out of SM fields suppressed by powers of a  high-energy scale $\Lambda$.  The leading operators which affect the Higgs production and decay amplitudes arise at dimension six, and so are suppressed by $1/\Lambda^2$.  Since no BSM states have been found so far, LHC results already indicate that the scale $\Lambda$ is higher than the scale $v= 246\,\text{GeV}$ of electroweak symmetry breaking. In a recent paper~\cite{Grojean:2013kd}, we studied a subset of these dimension-six operators
which modify the $h \to \gamma \gamma$ and $h \to Z \gamma $ decay rates, and  calculated the renormalization group evolution  of these operators, including operator mixing.

In this paper, we extend our previous RG analysis~\cite{Grojean:2013kd} to all dimension-six operators. We also compute the full contribution of the 59  dimension-six operators to the running of the usual  dimension $D \le 4$ operator coefficients of the SM Lagrangian.
These SM parameters  have anomalous dimension contributions of order $v^2/\Lambda^2$ (or equivalently, $m_H^2/\Lambda^2$) from  coefficients in the dimension-six Lagrangian. These terms correct the SM amplitudes at order $m_H^2/\Lambda^2$, which is the same order as the corrections from  dimension-six operators.

The set of independent higher dimensional operators
involving SM fields is given in  Ref.~\cite{Grzadkowski:2010es}, which showed that there are 59 independent dimension-six operators (assuming the conservation of baryon number), and reduced the set of operators from those of the earlier work  Ref.~\cite{Buchmuller:1985jz} by using the classical equations of motion to eliminate a few redundant operators. The choice of operator basis is not unique, and we will use the basis of
Ref.~\cite{Grzadkowski:2010es}, summarized in Table~\ref{op59}. The anomalous dimension matrix is a $59 \times 59$ matrix with $3481$ entries, not including flavor indices. Although some of the entries vanish due to the structure of the one-loop diagrams, most elements are non-zero. The $59$ operators can be grouped into eight classes defined in the next section. Our previous calculation~\cite{Grojean:2013kd} computed the $8 \times 8$ submatrix $\gamma_{44}$ of the $59 \times 59$ matrix.
%
%
\begin{table}
\begin{center}
\small
\begin{minipage}[t]{4.45cm}
\renewcommand{\arraystretch}{1.5}
\begin{tabular}[t]{c|c}
\multicolumn{2}{c}{$1:X^3$} \\
\hline
$Q_G$                & $f^{ABC} G_\mu^{A\nu} G_\nu^{B\rho} G_\rho^{C\mu} $ \\
$Q_{\widetilde G}$          & $f^{ABC} \widetilde G_\mu^{A\nu} G_\nu^{B\rho} G_\rho^{C\mu} $ \\
$Q_W$                & $\epsilon^{IJK} W_\mu^{I\nu} W_\nu^{J\rho} W_\rho^{K\mu}$ \\
$Q_{\widetilde W}$          & $\epsilon^{IJK} \widetilde W_\mu^{I\nu} W_\nu^{J\rho} W_\rho^{K\mu}$ \\
\end{tabular}
\end{minipage}
\begin{minipage}[t]{2.7cm}
\renewcommand{\arraystretch}{1.5}
\begin{tabular}[t]{c|c}
\multicolumn{2}{c}{$2:H^6$} \\
\hline
$Q_H$       & $(H^\dag H)^3$
\end{tabular}
\end{minipage}
\begin{minipage}[t]{5.1cm}
\renewcommand{\arraystretch}{1.5}
\begin{tabular}[t]{c|c}
\multicolumn{2}{c}{$3:H^4 D^2$} \\
\hline
$Q_{H\Box}$ & $(H^\dag H)\Box(H^\dag H)$ \\
$Q_{H D}$   & $\ \left(H^\dag D_\mu H\right)^* \left(H^\dag D_\mu H\right)$
\end{tabular}
\end{minipage}
\begin{minipage}[t]{2.7cm}

\renewcommand{\arraystretch}{1.5}
\begin{tabular}[t]{c|c}
\multicolumn{2}{c}{$5: \psi^2H^3 + \hbox{h.c.}$} \\
\hline
$Q_{eH}$           & $(H^\dag H)(\bar l_p e_r H)$ \\
$Q_{uH}$          & $(H^\dag H)(\bar q_p u_r \widetilde H )$ \\
$Q_{dH}$           & $(H^\dag H)(\bar q_p d_r H)$\\
\end{tabular}
\end{minipage}

\vspace{0.25cm}

\begin{minipage}[t]{4.7cm}
\renewcommand{\arraystretch}{1.5}
\begin{tabular}[t]{c|c}
\multicolumn{2}{c}{$4:X^2H^2$} \\
\hline
$Q_{H G}$     & $H^\dag H\, G^A_{\mu\nu} G^{A\mu\nu}$ \\
$Q_{H\widetilde G}$         & $H^\dag H\, \widetilde G^A_{\mu\nu} G^{A\mu\nu}$ \\
$Q_{H W}$     & $H^\dag H\, W^I_{\mu\nu} W^{I\mu\nu}$ \\
$Q_{H\widetilde W}$         & $H^\dag H\, \widetilde W^I_{\mu\nu} W^{I\mu\nu}$ \\
$Q_{H B}$     & $ H^\dag H\, B_{\mu\nu} B^{\mu\nu}$ \\
$Q_{H\widetilde B}$         & $H^\dag H\, \widetilde B_{\mu\nu} B^{\mu\nu}$ \\
$Q_{H WB}$     & $ H^\dag \tau^I H\, W^I_{\mu\nu} B^{\mu\nu}$ \\
$Q_{H\widetilde W B}$         & $H^\dag \tau^I H\, \widetilde W^I_{\mu\nu} B^{\mu\nu}$
\end{tabular}
\end{minipage}
\begin{minipage}[t]{5.2cm}
\renewcommand{\arraystretch}{1.5}
\begin{tabular}[t]{c|c}
\multicolumn{2}{c}{$6:\psi^2 XH+\hbox{h.c.}$} \\
\hline
$Q_{eW}$      & $(\bar l_p \sigma^{\mu\nu} e_r) \tau^I H W_{\mu\nu}^I$ \\
$Q_{eB}$        & $(\bar l_p \sigma^{\mu\nu} e_r) H B_{\mu\nu}$ \\
$Q_{uG}$        & $(\bar q_p \sigma^{\mu\nu} T^A u_r) \widetilde H \, G_{\mu\nu}^A$ \\
$Q_{uW}$        & $(\bar q_p \sigma^{\mu\nu} u_r) \tau^I \widetilde H \, W_{\mu\nu}^I$ \\
$Q_{uB}$        & $(\bar q_p \sigma^{\mu\nu} u_r) \widetilde H \, B_{\mu\nu}$ \\
$Q_{dG}$        & $(\bar q_p \sigma^{\mu\nu} T^A d_r) H\, G_{\mu\nu}^A$ \\
$Q_{dW}$         & $(\bar q_p \sigma^{\mu\nu} d_r) \tau^I H\, W_{\mu\nu}^I$ \\
$Q_{dB}$        & $(\bar q_p \sigma^{\mu\nu} d_r) H\, B_{\mu\nu}$
\end{tabular}
\end{minipage}
\begin{minipage}[t]{5.4cm}
\renewcommand{\arraystretch}{1.5}
\begin{tabular}[t]{c|c}
\multicolumn{2}{c}{$7:\psi^2H^2 D$} \\
\hline
$Q_{H l}^{(1)}$      & $(H^\dag i\overleftrightarrow{D}_\mu H)(\bar l_p \gamma^\mu l_r)$\\
$Q_{H l}^{(3)}$      & $(H^\dag i\overleftrightarrow{D}^I_\mu H)(\bar l_p \tau^I \gamma^\mu l_r)$\\
$Q_{H e}$            & $(H^\dag i\overleftrightarrow{D}_\mu H)(\bar e_p \gamma^\mu e_r)$\\
$Q_{H q}^{(1)}$      & $(H^\dag i\overleftrightarrow{D}_\mu H)(\bar q_p \gamma^\mu q_r)$\\
$Q_{H q}^{(3)}$      & $(H^\dag i\overleftrightarrow{D}^I_\mu H)(\bar q_p \tau^I \gamma^\mu q_r)$\\
$Q_{H u}$            & $(H^\dag i\overleftrightarrow{D}_\mu H)(\bar u_p \gamma^\mu u_r)$\\
$Q_{H d}$            & $(H^\dag i\overleftrightarrow{D}_\mu H)(\bar d_p \gamma^\mu d_r)$\\
$Q_{H u d}$ + h.c.   & $i(\widetilde H ^\dag D_\mu H)(\bar u_p \gamma^\mu d_r)$\\
\end{tabular}
\end{minipage}

\vspace{0.25cm}

\begin{minipage}[t]{4.75cm}
\renewcommand{\arraystretch}{1.5}
\begin{tabular}[t]{c|c}
\multicolumn{2}{c}{$8:(\bar LL)(\bar LL)$} \\
\hline
$Q_{ll}$        & $(\bar l_p \gamma_\mu l_r)(\bar l_s \gamma^\mu l_t)$ \\
$Q_{qq}^{(1)}$  & $(\bar q_p \gamma_\mu q_r)(\bar q_s \gamma^\mu q_t)$ \\
$Q_{qq}^{(3)}$  & $(\bar q_p \gamma_\mu \tau^I q_r)(\bar q_s \gamma^\mu \tau^I q_t)$ \\
$Q_{lq}^{(1)}$                & $(\bar l_p \gamma_\mu l_r)(\bar q_s \gamma^\mu q_t)$ \\
$Q_{lq}^{(3)}$                & $(\bar l_p \gamma_\mu \tau^I l_r)(\bar q_s \gamma^\mu \tau^I q_t)$
\end{tabular}
\end{minipage}
\begin{minipage}[t]{5.25cm}
\renewcommand{\arraystretch}{1.5}
\begin{tabular}[t]{c|c}
\multicolumn{2}{c}{$8:(\bar RR)(\bar RR)$} \\
\hline
$Q_{ee}$               & $(\bar e_p \gamma_\mu e_r)(\bar e_s \gamma^\mu e_t)$ \\
$Q_{uu}$        & $(\bar u_p \gamma_\mu u_r)(\bar u_s \gamma^\mu u_t)$ \\
$Q_{dd}$        & $(\bar d_p \gamma_\mu d_r)(\bar d_s \gamma^\mu d_t)$ \\
$Q_{eu}$                      & $(\bar e_p \gamma_\mu e_r)(\bar u_s \gamma^\mu u_t)$ \\
$Q_{ed}$                      & $(\bar e_p \gamma_\mu e_r)(\bar d_s\gamma^\mu d_t)$ \\
$Q_{ud}^{(1)}$                & $(\bar u_p \gamma_\mu u_r)(\bar d_s \gamma^\mu d_t)$ \\
$Q_{ud}^{(8)}$                & $(\bar u_p \gamma_\mu T^A u_r)(\bar d_s \gamma^\mu T^A d_t)$ \\
\end{tabular}
\end{minipage}
\begin{minipage}[t]{4.75cm}
\renewcommand{\arraystretch}{1.5}
\begin{tabular}[t]{c|c}
\multicolumn{2}{c}{$8:(\bar LL)(\bar RR)$} \\
\hline
$Q_{le}$               & $(\bar l_p \gamma_\mu l_r)(\bar e_s \gamma^\mu e_t)$ \\
$Q_{lu}$               & $(\bar l_p \gamma_\mu l_r)(\bar u_s \gamma^\mu u_t)$ \\
$Q_{ld}$               & $(\bar l_p \gamma_\mu l_r)(\bar d_s \gamma^\mu d_t)$ \\
$Q_{qe}$               & $(\bar q_p \gamma_\mu q_r)(\bar e_s \gamma^\mu e_t)$ \\
$Q_{qu}^{(1)}$         & $(\bar q_p \gamma_\mu q_r)(\bar u_s \gamma^\mu u_t)$ \\
$Q_{qu}^{(8)}$         & $(\bar q_p \gamma_\mu T^A q_r)(\bar u_s \gamma^\mu T^A u_t)$ \\
$Q_{qd}^{(1)}$ & $(\bar q_p \gamma_\mu q_r)(\bar d_s \gamma^\mu d_t)$ \\
$Q_{qd}^{(8)}$ & $(\bar q_p \gamma_\mu T^A q_r)(\bar d_s \gamma^\mu T^A d_t)$\\
\end{tabular}
\end{minipage}

\vspace{0.25cm}

\begin{minipage}[t]{3.75cm}
\renewcommand{\arraystretch}{1.5}
\begin{tabular}[t]{c|c}
\multicolumn{2}{c}{$8:(\bar LR)(\bar RL)+\hbox{h.c.}$} \\
\hline
$Q_{ledq}$ & $(\bar l_p^j e_r)(\bar d_s q_{tj})$
\end{tabular}
\end{minipage}
\begin{minipage}[t]{5.5cm}
\renewcommand{\arraystretch}{1.5}
\begin{tabular}[t]{c|c}
\multicolumn{2}{c}{$8:(\bar LR)(\bar L R)+\hbox{h.c.}$} \\
\hline
$Q_{quqd}^{(1)}$ & $(\bar q_p^j u_r) \epsilon_{jk} (\bar q_s^k d_t)$ \\
$Q_{quqd}^{(8)}$ & $(\bar q_p^j T^A u_r) \epsilon_{jk} (\bar q_s^k T^A d_t)$ \\
$Q_{lequ}^{(1)}$ & $(\bar l_p^j e_r) \epsilon_{jk} (\bar q_s^k u_t)$ \\
$Q_{lequ}^{(3)}$ & $(\bar l_p^j \sigma_{\mu\nu} e_r) \epsilon_{jk} (\bar q_s^k \sigma^{\mu\nu} u_t)$
\end{tabular}
\end{minipage}
\end{center}
\caption{\label{op59}
The 59 independent dimension-six operators built from Standard Model fields which conserve baryon number, as given in Ref.~\cite{Grzadkowski:2010es}. The operators are divided into eight classes: $X^3$, $H^6$, etc. Operators with $+\hbox{h.c.}$ in the table heading also have hermitian conjugates, as does the $\psi^2H^2D$ operator $Q_{Hud}$. The subscripts $p,r,s,t$ are flavor indices, The notation is described in Sec.~\ref{sec:lag}.
}
\end{table}
%
%

The full $59\times 59$ matrix is lengthy, and we give partial results here. Ref.~\cite{Grojean:2013kd} found that the $\lambda$ and Yukawa coupling terms were numerically more important than the gauge terms. In this paper, we give the $\lambda$ , $\lambda^2$, $\lambda y^2$ one-loop contributions to the anomalous dimension, which gives the full $\lambda$ dependence in the limit of vanishing gauge coupling. There are large combinatorial factors $\sim 100$ in some of the terms.

There are terms in the anomalous dimension matrix of order 1. These arise from diagrams involving external gauge fields, and are order 1 because gauge couplings are absorbed into the gauge field-strengths in our counting scheme, defined in Sec.~\ref{sec:adim}. We give one example of such a contribution at the end of Sec.~\ref{sec:adim}, which gives mixing between ``tree'' and ``loop'' operators, discussed in
Refs.~\cite{Arzt:1994gp,Elias-Miro:2013gya,Jenkins:2013fya,Manohar:2013rga}.

The outline of the paper is as follows:
In Sec.~\ref{sec:lag}, we summarize the Lagrangian we use,  our notational conventions, and the SM equations of motion. A review of well-known results on renormalization and the equations of motion is given in Sec.~\ref{sec:eom}.
The dimension-six contribution to the SM RGE is given in Sec.~\ref{sec:smrge}. The structure of the $59\times 59$ anomalous dimension matrix, our power counting scheme, and the terms we present in this paper are given in Sec.~\ref{sec:adim}. The dimension-six RGE equations are given in Sec.~\ref{sec:lam}.

Calculations are done in the \msbar\ scheme using dimensional regularization in $d=4-2\epsilon$ dimensions in background field gauge. The anomalous dimensions of gauge invariant operators do not depend on a choice of gauge, and so are the same in the broken and unbroken theory in the \msbar\ scheme.

\section{The Lagrangian and Equations of Motion}\label{sec:lag}

\subsection{The Lagrangian}

The Lagrangian we use is given by $\mathcal{L}=\mathcal{L}_{\rm SM}+\lsix$, the sum of the SM Lagrangian
\begin{align}
\mathcal{L} _{\rm SM} &= -\frac14 G_{\mu \nu}^A G^{A\mu \nu}-\frac14 W_{\mu \nu}^I W^{I \mu \nu} -\frac14 B_{\mu \nu} B^{\mu \nu}
+ (D_\mu H^\dagger)(D^\mu H)
+\sum_{\psi=q,u,d,l,e} \overline \psi\, i \slashed{D} \, \psi\nn
&-\lambda \left(H^\dagger H -\frac12 v^2\right)^2- \biggl[ H^{\dagger j} \overline d\, Y_d\, q_{j} + \widetilde H^{\dagger j} \overline u\, Y_u\, q_{j} + H^{\dagger j} \overline e\, Y_e\,  l_{j} + \hbox{h.c.}\biggr]
\label{sm}
\end{align}
and the dimension-six Lagrangian, which is given schematically by
\begin{align}
\lsix &= \sum_i C_i \, Q_i \,.
\label{six}
\end{align}
$H$ is an $SU(2)$ scalar doublet with hypercharge $\hyp_H=1/2$. With this normalization convention,
the Higgs boson mass is $m_H^2=2\lambda v^2$, with $v \sim 246$~GeV and  the fermion mass matrices are $M_{u,d,e}=Y_{u,d,e}\, v /\sqrt 2$.

The gauge covariant derivative is $D_\mu = \partial_\mu + i g_3 T^A A^A_\mu + i g_2  t^I W^I_\mu + i g_1 \hyp B_\mu$, where $T^A$ are the $SU(3)$ generators,  $t^I=\tau^I/2$ are the $SU(2)$ generators, and $\hyp$ is the $U(1)$ hypercharge generator.  $SU(2)$ indices $j,k$ and $I,J,K$ are in the fundamental and adjoint representations, respectively, and $SU(3)$ indices $A,B,C$ are in the adjoint representation.
$\widetilde H$ is defined by
\begin{align}
\widetilde H_j &= \epsilon_{jk} H^{\dagger\, k}
\end{align}
where the $SU(2)$ invariant tensor $\epsilon_{jk}$ is defined by $\epsilon_{12}=1$ and $\epsilon_{jk}=-\epsilon_{kj}$, $j,k=1,2$.  Fermion fields $q$ and $l$ are left-handed fields, and $u$, $d$ and $e$ are right-handed fields.

We have suppressed flavor indices in Eq.~(\ref{sm}). All fermion fields have a flavor index $p=1,2,3$ for the three generations, and the Yukawa matrices $Y_{u,d,e}$ are matrices in flavor space. Explicitly,
\begin{align}
H^{\dagger j} \overline d\, Y_d\, q_{j} &= H^{\dagger j} \overline d_p\, [Y_d]_{pr}\, q_{rj}
\end{align}
and similarly for the other terms. Flavor indices are denoted by $p,r,s,t$.
We  work in the weak eigenstate basis, with $u_i=\left\{u_R,c_R,t_R\right\}$,
$d_i=\left\{d_R,s_R,b_R\right\}$, and
\begin{align}
q_1 &= \left[ \begin{array}{c} u_L \\ d_L^\prime \end{array}\right], &
q_2 &= \left[ \begin{array}{c} c_L \\ s_L^\prime \end{array}\right], &
q_3 &= \left[ \begin{array}{c} t_L \\ b_L^\prime \end{array}\right], &
\left[ \begin{array}{c} d_L^\prime \\ s_L^\prime \\ b_L^\prime \end{array}\right] &=
V_{\rm CKM} \, \left[ \begin{array}{c} d_L \\ s_L \\ b_L \end{array}\right] ,
\end{align}
where $V_{\rm CKM}$ is the quark mixing matrix.

The coefficients $C_i$ of the dimension-six Lagrangian have mass dimension $-2$. The sum on $i$ in Eq.~(\ref{six}) is over the 59 operators in Table~\ref{op59}.
The only (notational) change from Ref.~\cite{Grzadkowski:2010es} is the replacement of $\varphi$ by $H$ for the Higgs field. Note that $Q_{uH}$ and $Q_{Hu}$, etc.\ are different operators. We use the convention $\widetilde F_{\mu \nu} =(1/2) \epsilon_{\mu \nu \alpha \beta} F^{\alpha \beta}$ with $\epsilon_{0123}=+1$.
The operators are divided into eight classes, which are denoted by $1:X^3$, $2:H^6$, $3:H^4 D^2$, $4:X^2 H^2$, $5:\psi^2H^3$, $6:\psi^2 H X$, $7:\psi^2 H^2 D$, and $8:\psi^4$ in terms of the field content and number of covariant derivatives, with  $X$ denoting a gauge field strength tensor. We will use this schematic notation for other operators that occur in our analysis. For example, the penguin operator $\overline q T^A \gamma^\mu q \left[D^\nu,G_{\nu \mu}\right]^A$ is a $\psi^2 X D$ operator.

The coefficients $C_i$ are then divided into eight blocks, $i=1,\ldots,8$, with block 1 containing four coefficients for the $X^3$ operators, etc.
The anomalous dimension matrix also breaks up into blocks with $\gamma_{14}$ denoting the $4 \times 8$ submatrix in the $X^3 - X^2 H^2$ sector, etc. The notation in Table~\ref{op59} suppresses flavor indices. Two sample terms in Eq.~(\ref{six}) including flavor indices are
\begin{align}
C_{\substack{eu \\ prst}} Q_{\substack{eu \\ prst}}+\left[ C_{\substack{ledq \\ prst}} Q_{\substack{ledq \\ prst}} + \hbox{h.c.}\right]
\end{align}
where the hermitian conjugate is added for non-self-conjugate operators. The coefficients of the self-conjugate operators are hermitian tensors, so that
\begin{align}
C^*_{\substack{eu \\ prst}} &= C^*_{\substack{eu \\ rpts}}\,,
\end{align}
and similarly for the other coefficients.

\subsection{SM Equations of Motion}

The SM equations of motion  play an important role in the following analysis, so we summarize them here.
The SM equations of motion from Eq.~(\ref{sm}) are
\begin{align}
D^2 H_k -\lambda v^2 H_k +2 \lambda (H^\dagger H) H_k + \overline q^j\, Y_u^\dagger\, u \epsilon_{jk} + \overline d\, Y_d\, q_k +\overline e\, Y_e\,  l_k
&=0 \,,
\label{eomH}
\end{align}
for the Higgs field,
\begin{align}
i\slashed{D}\, q_j &= Y_u^\dagger\, u\, \widetilde H_j + Y_d^\dagger\, d\, H_j \,, &
i\slashed{D}\,  d &= Y_d\,  q_j\, H^{\dagger\, j} \,, &
i\slashed{D}\, u &= Y_u\, q_j\, \widetilde H^{\dagger\, j}\,, \nn
i\slashed{D} \, l_j &= Y_e^\dagger\, e  H_j  \,, &
i\slashed{D} \, e &= Y_e\, l_j H^{\dagger\, j}\,,
\label{eompsi}
\end{align}
for the fermion fields, and
\begin{align}
\left[D^\alpha , G_{\alpha \beta} \right]^A &= g_3  j_\beta^A, &
\left[D^\alpha , W_{\alpha \beta} \right]^I &= g_2  j_\beta^I, &
D^\alpha B_{\alpha \beta} &= g_1  j_\beta ,
\label{eomX}
\end{align}
for the gauge fields, where $\left[D^\alpha , F_{\alpha \beta} \right]$ is the covariant derivative in the adjoint representation. The gauge currents are
\begin{align}
j_\beta^A &=\sum_{\psi=u,d,q} \overline \psi \, T^A \gamma_\beta  \psi\,,\nn
j_\beta^I &= \frac 12 \overline q \, \tau^I \gamma_\beta  q + \frac12 \overline l \, \tau^I \gamma_\beta  l + \frac12 H^\dagger \, i\overleftrightarrow D_\beta^I H\,, \nn
j_\beta &=\sum_{\psi=u,d,q,e,l} \overline \psi \, \hyp_i \gamma_\beta  \psi + \frac12 H^\dagger \, i\overleftrightarrow D_\beta H\,,
\end{align}
where $\hyp_i$ are the $U(1)$ hypercharges of the fermions, and
\begin{align}
H^\dagger \, i\overleftrightarrow D_\beta H &= i H^\dagger (D_\beta H) - i (D_\beta H^\dagger) H \,,\nn
H^\dagger \, i\overleftrightarrow D_\beta^I H &= i H^\dagger \tau^I (D_\beta H) - i (D_\beta H^\dagger)\tau^I H\,.
\end{align}

\section{Operator Renormalization and the Equations of Motion}\label{sec:eom}

In this section, we review some well-known results about equations of motion (EOM) and renormalization in field theory.
One can make field redefinitions on the Lagrangian, which is a change of variables in a path integral, and so does not affect $S$-matrix elements~\cite{Politzer:1980me}.\footnote{Field redefinitions can affect Green's functions, since the source terms get modified.}
Field redefinitions can be systematically used to eliminate redundant operators from the Lagrangian. In our case, $\mathcal{L}=\mathcal{L}_{\rm SM}+ \lsix$, so a small field redefinition of order $1/\Lambda^2$ can be used to shift $\lsix$ by operators proportional to the  \emph{classical} EOM  from the SM Lagrangian. For example, the dimension-six operator
\begin{align}
E_{H \Box} &= [H^\dagger H] [H^\dagger (D^2 H) + (D^2 H^\dagger) H]
\label{ebox}
\end{align}
can be converted to
\begin{align}
\widetilde E_{H \Box} & = 2\lambda v^2 (H^\dagger H)^2 -4 \lambda Q_H -\left( [Y_u^\dagger]_{rs}\, Q_{\substack{u H \\ rs }} + [Y_d^\dagger]_{rs} \, Q_{\substack{d H \\ rs }}+ [Y_e^\dagger]_{rs}\, Q_{\substack{eH \\ rs }}+ \hbox{h.c.} \right)
\label{ebox2}
\end{align}
Explicitly,
\begin{align}
\mathcal{L}_{\rm SM} + \frac{c}{\Lambda^2} E_{H \Box} \to \mathcal{L}_{\rm SM} + \frac{c}{\Lambda^2} \widetilde E_{H \Box}
+ \mathcal{O}\left( \frac{1}{\Lambda^4} \right)
\end{align}
by the field redefinition
\begin{align}
H \to H + \frac{c}{\Lambda^2} (H^\dagger H) H\,
\end{align}
which is equivalent to using Eq.~(\ref{eomH}) to convert $E_{H \Box}$ to $\widetilde E_{H \Box}$.
Thus, to first order in $1/\Lambda^2$, we can eliminate dimension-six EOM operators. At higher orders in $1/\Lambda^2$, it is necessary to systematically use field redefinitions to eliminate redundant operators~\cite{Georgi:1991ch,Manohar:1997qy,Manohar:1996cq,Einhorn:2001kj}.

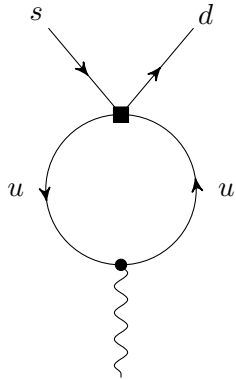
\begin{figure}
\centering
\begin{tikzpicture}[
decoration={
	markings,
	mark=at position 0.55 with {\arrow[scale=1.5]{stealth'}};
}]

\draw[postaction=decorate] (90:1) arc (90:270:1) ;
\draw[postaction=decorate] (-90:1) arc (-90:90:1) ;

\filldraw (-0.1,0.9) rectangle (0.1,1.1);

\filldraw (270:1) circle (0.075);

\draw[postaction=decorate] (90:1) -- +(50:1.5) ;
\draw[postaction=decorate] (90:1)+(130:1.5) -- (90:1) ;

\draw[decorate,decoration={snake}]  (270:1) -- +(270:1.5);

\draw (90:1)+(130:1.75) node [align=center] {$s$} ;
\draw (90:1)+(50:1.75) node [align=center] {$d$} ;
\draw (180:1.4) node [align=center] {$u$} ;
\draw (0:1.4) node [align=center] {$u$} ;

\end{tikzpicture}
\caption{\label{fig:penguin}
Penguin diagram contributing to $s \to d$ transitions.}
\end{figure}

The counterterms generated by one-loop graphs from $\lsix$ need not be in the standard basis chosen for the dimension-six operators. It is necessary to first compute the renormalization counterterms, and then convert them to the standard basis using a field redefinition.
A famous example of this procedure is the renormalization of the effective Lagrangian for weak decays~\cite{Gilman:1979bc,Georgi:1985kw}. One can use an operator basis involving only four-quark operators, such as
\begin{align}
O_q &=\overline u\, \gamma^\mu P_L\, s\ \overline d\, \gamma_\mu P_L\, u
\label{4quark}
\end{align}
for $s \to d$ transitions. However, the penguin graph Fig.~\ref{fig:penguin}
requires a counterterm proportional to
\begin{align}
O_P &= \overline d\, T^A \gamma^\mu P_L\, s\ g_3\left[D^\nu, G_{\nu \mu}\right]^A\,.
\label{penguin}
\end{align}
The standard procedure used is to convert this back to a four-quark operator using the gauge field equation of motion Eq.~(\ref{eomX}),
\begin{align}
O_P &\to \overline d\, T^A \gamma^\mu P_L\, s \
\sum_q g_3^2\left[ \overline q\, T^A \gamma^\mu P_L\, q + \overline q\, T^A \gamma^\mu P_R\, q\right]\,,
\end{align}
so that one can study the anomalous dimension matrix in the basis of four-quark operators.

In general, let $E_i$ be the dimension-six EOM operators generated by field redefinitions on the SM Lagragian, so that $E_i=0$ by the classical SM equations of motion. Then the general dimension-six Lagrangian is
\begin{align}
\lsix &= \sum_{i=1}^{59} C_i Q_i + \sum_r D_r E_r
\label{2.16}
\end{align}
including redundant EOM operators. The RGE has the form
\begin{align}
\mu \frac{\rd}{\rd \mu} \left[ \begin{array}{c} Q_i \\ E_r \end{array}\right]
&= \left[ \begin{array}{cc} -\gamma_{ji} & -a_{si} \\ 0 & -b_{sr} \end{array}\right] \left[ \begin{array}{c} Q_j \\ E_s \end{array}\right]\,.
\label{2.17}
\end{align}
The lower left block of this matrix  vanishes, since the EOM operators are renormalized among themselves~\cite{Politzer:1980me}. The operators $E_r$ do not contribute to $S$-matrix elements, so their $\mu$ derivative cannot contain $Q_i$ which have non-zero $S$-matrix elements.
Eq.~(\ref{2.17}) implies that the anomalous dimension matrix for the coefficients has the form
\begin{align}
\mu \frac{\rd}{\rd \mu} \left[ \begin{array}{c} C_i \\ D_r \end{array}\right]
&= \left[ \begin{array}{cc} \gamma_{ij} & 0 \\ a_{rj} & b_{rs} \end{array}\right] \left[ \begin{array}{c} C_j \\ D_s \end{array}\right]\,.
\label{2.18}
\end{align}
The $E_i$ operators in Eq.~(\ref{2.16}) can be dropped for $S$-matrix element calculations, i.e.\ we only need the values of $C_i$.
From Eq.~(\ref{2.18}), we see that in this case, the RGE reduces to
\begin{align}
\mu \frac{\rd}{\rd \mu}C_i&= \gamma_{ij} C_j \,,
\label{2.19}
\end{align}
with no reference to the EOM operators.

It is important to remember that this conclusion does not mean that EOM operators do not enter the calculation. The penguin operator counterterm Eq.~(\ref{penguin}) is written as
\begin{align}
O_P &= \overline d\, T^A \gamma^\mu P_L\, s \
\sum_q g_3^2\left[ \overline q\, T^A \gamma^\mu P_L\, q + \overline q\, T^A \gamma^\mu P_R\, q\right] + E\,, \nn
E &= \overline d\, T^A \gamma^\mu P_L\, s\ g_3\left[D^\nu, G_{\nu \mu}\right]^A - \overline d\, T^A \gamma^\mu P_L\, s \
\sum_q g_3^2\left[ \overline q\, T^A \gamma^\mu P_L\, q + \overline q\, T^A \gamma^\mu P_R\, q\right]
\end{align}
where $E=0$ is an EOM operator, which can be dropped. The remaining four-quark contribution  enters the RGE for the four-quark operators.

There is an important consequence of the above analysis, which has led to some confusion in the literature. One cannot  identify the structure of the
anomalous dimension matrix simply from one-particle irreducible one-loop diagrams when the EOM are used to reduce operators to a standard basis.
For example, the penguin operator is a $\psi^2 X D$ operator, but leads to a $\psi^4$ counterterm after using the equations of motion.
This is because the EOM can generate new operators and mixing for which \emph{no irreducible graphs exist.} This subtlety does not affect the $\gamma_{44}$ anomalous dimension given in Ref.~\cite{Grojean:2013kd}.

In some cases, authors have used a redundant basis of operators, i.e.\ an overcomplete set of operators in which some operators can be eliminated using the equations of motion. An example of such a procedure is to include both the penguin operator Eq.~(\ref{penguin}) and the various four-quark operators it generates, such as those on the r.h.s.\ of  Eq.~(\ref{2.18}). Schematically, assume that  the theory has operators
$O_{1,2,3}$, the EOM is $O_2=O_3$, and that the RGE after eliminating the redundant operator $O_3$ is
\setlength{\arraycolsep}{0.2cm}
\renewcommand{\arraystretch}{1.2}
\begin{align}
\mu \frac{\rd}{\rd \mu} \left[ \begin{array}{c}O_1 \\ O_2 \end{array}\right]
&= - \left[ \begin{array}{cc} \gamma_{11} & \gamma_{21}  \\  \gamma_{12}  & \gamma_{22}  \end{array}\right] \left[ \begin{array}{c} O_1 \\ O_2 \end{array}\right]\,.
\label{2.24}
\end{align}
The RGE including the redundant operator $O_3$ has the form
\begin{align}
\mu \frac{\rd}{\rd \mu} \left[ \begin{array}{c}O_1 \\ O_2 \\ O_3 \end{array}\right]
&= - \left[ \begin{array}{ccc} \gamma_{11} & \gamma_{21} +a _1 & -a_1 \\  \gamma_{12}  & \gamma_{22} +a _2 & -a _2  \\
0 & a_3 & -a_3
\end{array}\right] \left[ \begin{array}{c} O_1 \\ O_2  \\ O_3\end{array}\right]\,.
\label{2.25}
\end{align}
with $a_{1,2,3}$ arbitrary. In this case, the anomalous dimension matrix is not uniquely determined, since one can always add linear combinations of EOM operators to the RGE by making field redefinitions. The parameters $a_i$ depend on the gauge and renormalization scheme, since different choices can differ implicitly by field redefinitions. Note that even the $2\times 2$ submatrix in the $O_{1,2}$ sector is not unique.

\section{Running of SM terms due to $\lsix$}\label{sec:smrge}

The SM coefficients have contributions from $\lsix$ proportional to $v^2$, or equivalently, $m_H^2$. The existence of such terms is not surprising. In the usual analysis of $K^0 - \overline K^0$ mixing due to $\Delta S=2$ weak interactions,  the $\Delta S=2$ Lagrangian
\begin{align}
L^{(\Delta S=2)} &= C_2\ \overline d \, \gamma^\mu P_L \, s\ \overline d \, \gamma_\mu P_L \, s
\end{align}
has terms in the RGE of the form~\cite{Georgi:1985kw}
\begin{align}
\mu \frac{\rd}{\rd \mu} C_2 &\propto m_q^2\, C_q^2
\end{align}
where $C_q$ are the coefficients of the $\Delta S=1$ four-quark operators such as Eq.~(\ref{4quark}), and $m_q$ is a quark mass. Mass terms in the EFT can compensate for powers of $1/M_W$, since particle masses can appear in the numerator of divergent terms when dimensional regularization is used. In the case of SM running from $\lsix$, the only dimensionful parameter in the SM that can appear in the numerator is the Higgs vacuum expectation value $v$, or equivalently, the Higgs mass $m_H^2=2\lambda v^2$.

We list here the full one-loop contributions to the SM RGE from $\lsix$. These terms are \emph{in addition} to the usual SM anomalous dimensions.
\begin{align}
\mu \frac{\rd}{\rd \mu} \lambda &= \frac{m_H^2 }{16\pi^2}  \biggl[ 12 C_H+ \left(- 32 \lambda + \frac{10}3 g_2^2   \right) C_{H \Box} +\left(12 \lambda -\frac32 g_2^2 + 6 g_1^2  \hyp_H^2  \right) C_{H D}+2 \eta_1 +2 \eta_2  \nn
&+12 g_2^2 c_{F,2} C_{HW} + 12 g_1^2 \hyp_H^2 C_{HB} + 6 g_1 g_2 \hyp_H C_{HWB}  +\frac43 g_2^2 C^{(3)}_{\substack{H l \\ tt}} +\frac43 g_2^2 N_c C^{(3)}_{\substack{H q \\ tt}}
\biggr]\,, \nn
\mu \frac{\rd}{\rd \mu}   m_H^2  &= \frac{m_H^4}{16\pi^2}  \left[ - 4  C_{H \Box} + 2  C_{H D} \right] \,,\nn
\mu \frac{\rd}{\rd \mu}  [Y_u]_{rs} &=  \frac{m_H^2 }{16\pi^2}  \biggl[3 C_{\substack{uH \\ sr}}^* - C_{H \Box} [Y_u]_{rs} + \frac12 C_{H D} [Y_u]_{rs} -
[Y_u]_{rt} \left( C^{(1)}_{\substack{H q \\ ts}} - 3 C^{(3)}_{\substack{H q \\ ts}} \right)+ C_{\substack{H u \\ rt}} [Y_u]_{ts} \nn
&-C_{\substack{H ud \\ rt}}  [Y_d]_{ts}  -2 \left( C^{(1)}_{\substack{qu \\ psrt}} + c_{F,3} C^{(8)}_{\substack{qu \\ psrt}} \right) [Y_u]_{tp}
- C^{(1)*}_{\substack{lequ \\ ptsr}} [Y_e^\dagger]_{pt}+ N_c C^{(1)*}_{\substack{quqd \\ srpt}} [Y_d^\dagger]_{pt}\nn
&+\frac12 \left( C^{(1)*}_{\substack{quqd \\ prst}} + c_{F,3} C^{(8)*}_{\substack{quqd \\ prst}}  \right) [Y_d^\dagger]_{pt}
  \biggr]\,, \nn
 \mu \frac{\rd}{\rd \mu}  [Y_d]_{rs} &=  \frac{m_H^2 }{16 \pi^2} \biggl[3 C_{\substack{dH \\ sr}}^*  - C_{H \Box} [Y_d]_{rs} + \frac12 C_{H D} [Y_d]_{rs} + [Y_d]_{rt} \left( C^{(1)}_{\substack{H q \\ ts}} +3  C^{(3)}_{\substack{H q \\ ts}}\right)  - C_{\substack{H d \\ rt}} [Y_d]_{ts} \nn
&- [Y_u]_{ts} C^*_{\substack{H ud \\ tr}}  -2 \left( C^{(1)}_{\substack{qd \\ psrt}} + c_{F,3} C^{(8)}_{\substack{qd \\ psrt}} \right) [Y_d]_{tp}
+C_{\substack{ledq \\ ptrs}}[Y_e]_{tp}  + N_c C^{(1)*}_{\substack{quqd \\ ptsr}} [Y_u^\dagger]_{pt}\nn
&+ \frac12 \left( C^{(1)*}_{\substack{quqd \\ sptr}} + c_{F,3} C^{(8)*}_{\substack{quqd \\ sptr}} \right) [Y_u^\dagger]_{tp} \biggr] \,,\nn
\mu \frac{\rd}{\rd \mu}  [Y_e]_{rs} &=  \frac{m_H^2 }{16\pi^2}  \biggl[3 C_{\substack{eH \\ sr}}^* - C_{H \Box} [Y_e]_{rs} + \frac12 C_{H D} [Y_e]_{rs} +
[Y_e]_{rt} \left( C^{(1)}_{\substack{H l \\ ts}} +3 C^{(3)}_{\substack{H l \\ ts}}\right) -C_{\substack{H e \\ rt}}  [Y_e]_{ts} \nn
&-2 C_{\substack{le \\ psrt}}[Y_e]_{tp} + N_c C^{*}_{\substack{ledq \\ srpt}}[Y_d]_{pt}
-N_c C^{(1)*}_{\substack{lequ \\ srpt}} [Y_u^\dagger]_{pt} \biggr]\,,
\label{smrge}
\end{align}
\begin{align}
\mu \frac{\rd g_3}{\rd \mu}   &=  -4\frac{m_H^2 }{16\pi^2} g_3 C_{H G} \,, &
\mu \frac{\rd g_2}{\rd \mu}   &=  -4\frac{m_H^2 }{16\pi^2} g_2 C_{H W}\,, &
\mu \frac{\rd g_1}{\rd \mu}   &=  -4\frac{m_H^2 }{16\pi^2} g_1 C_{H B}\,, \nn
\mu \frac{\rd}{\rd \mu}  \theta_3  &=  - \frac{4m_H^2}{g_3^2} C_{H\widetilde G}\,, &
\mu \frac{\rd}{\rd \mu}  \theta_2  &=  - \frac{4m_H^2}{g_2^2} C_{H\widetilde W}\,, &
\mu \frac{\rd}{\rd \mu}  \theta_1  &=  - \frac{4m_H^2}{g_1^2} C_{H\widetilde B}\,,
\label{smrgeA}
\end{align}
where
\begin{align}
\eta_1 &= \left( \frac12 N_c C_{\substack{dH \\ rs}} [Y_d]_{sr}
 +  \frac12 N_c C_{\substack{uH \\ rs}} [Y_u]_{sr}
+ \frac12 C_{\substack{eH \\ rs}} [Y_e]_{sr} \right) + h.c.\,, \nn
\eta_2 &=  -2 N_c C_{\substack{H q \\ rs}}^{(3)} [Y_u^\dagger   Y_u ]_{sr}
-2 N_c C_{\substack{H q \\ rs}}^{(3)} [Y_d^\dagger  Y_d  ] _{sr}
+ N_c C_{\substack{H ud \\ rs}}  [Y_d Y_u^\dagger   ]_{sr}+ N_c C^*_{\substack{H ud \\ rs}}  [Y_u Y_d^\dagger   ]_{rs} -2  C_{\substack{H l \\ rs}}^{(3)} [Y_e^\dagger  Y_e  ] _{sr}  \,,\nn
\label{etadef}
\end{align}
$N_c=3$ is the number of colors, $c_{F,3}=4/3$, and $c_{A,2}=2$.
The $\theta$ terms are normalized so that $L = (\theta g^2/32\pi^2) \widetilde F_{\mu \nu}^A F^{A\,\mu\nu}$ for each gauge group.\footnote{Transformations of $\theta_{1,2,3}$ under flavor transformations, and the basis invariant definition of $\overline\theta$ angles in the electroweak theory is discussed in Ref.~\cite{Jenkins:2009dy}.}

The form of Eq.~(\ref{smrge}) depends on the choice of basis for $\lsix$, since EOM have been used to eliminate redundant operators. One can see from Eq.~(\ref{ebox2}) that the EOM contain both dimension-four operators $(H^\dagger H)^2$, and dimension-six operators such as $Q_H$, so the dimension-four terms depend on the basis choice for the dimension-six terms.

Eq.~(\ref{smrge}) affects SM amplitudes at order $m_H^2/\Lambda^2$, and is just as important as the evolution of $\lsix$.
For $\Lambda \sim 1$~TeV, the terms in Eq.~(\ref{smrge}) are more important than two-loop contributions to the SM running. The stability of the Higgs scalar potential is very sensitive to the precise value of $\lambda$, so the $\lsix$ contribution will affect the relation between $m_H$ and the scalar self-coupling. Eq.~(\ref{smrge}) also plays a role in the evolution of Yukawa couplings. Retaining only the top-quark Yukawa coupling,
\begin{align}
Y_e &\to \left[ \begin{array}{ccc} 0 & 0 & 0 \\ 0 & 0 & 0 \\ 0 & 0 & 0 \end{array}\right], & Y_d & \to \left[ \begin{array}{ccc} 0 & 0 & 0 \\ 0 & 0 & 0 \\ 0 & 0 & 0 \end{array}\right], & Y_u & \to \left[ \begin{array}{ccc} 0 & 0 & 0 \\ 0 & 0 & 0 \\ 0 & 0 & y_t \end{array}\right],
\end{align}
gives from Eq.~(\ref{smrge}),
\begin{align}
\mu \frac{\rd}{\rd \mu}  [Y_d]_{rs} &=  \frac{m_H^2 }{16\pi^2}  \biggl[3 C_{\substack{dH \\ sr}}^*
- [Y_u]_{ts} C^*_{\substack{H ud \\ tr}} + N_c C^{(1)*}_{\substack{quqd \\ ptsr}} [Y_u^\dagger]_{pt}
+ \frac12 \left( C^{(1)*}_{\substack{quqd \\ sptr}} + c_{F,3} C^{(8)*}_{\substack{quqd \\ sptr}} \right) [Y_u^\dagger]_{tp} \biggr] \,,\nn
\mu \frac{\rd}{\rd \mu}  [Y_e]_{rs} &=  \frac{m_H^2 }{16\pi^2}  \biggl[3 C_{\substack{eH \\ sr}}^*-N_c C^{(1)*}_{\substack{lequ \\ srpt}} [Y_u^\dagger]_{pt} \biggr]\,.
\end{align}
In the SM, $\mu\, {\rd} Y_i/{\rd \mu} \propto Y_i$ so these higher dimension terms can be more important than the SM running, depending on
the flavor structure of $\lsix$.

\section{Structure of the Anomalous Dimension Matrix}\label{sec:adim}

The SM at energies above the electroweak scale is a weakly coupled gauge theory, and SM gauge boson interactions are proportional to the gauge boson coupling $g$. For this reason, it is useful to use rescaled operators $\widetilde Q_i$ with coefficients $\widetilde C_i$, including an explicit factor of the gauge coupling for each field strength tensor $X$, instead of the basis $Q_i$ in Table~\ref{op59}. Thus $\widetilde Q_G=g_3^3 Q_G$, etc. One can trivially convert between the two conventions. If
$\widetilde Q_i = \zeta_i Q_i$,
then the rescaled coefficients and anomalous dimensions are
\begin{align}
\widetilde C_i &= \zeta_i^{-1} C_i\,, & \widetilde \gamma_{ij} &= \zeta_i^{-1} \gamma_{ij} \zeta_j\,.
\label{3.3}
\end{align}

We first discuss the structure of the one-loop anomalous dimension matrix, where all gauge couplings are treated as order $g$ and all Yukawa couplings as order $y$. The $\psi^2 H^3$ and $\psi^2 X H$ operators have a single chirality flip.
 It is convenient to absorb a numerical factor of order $y$ into these operators for the purposes of the present discussion,\footnote{We do not include any factors of $y$ in the $(\overline L R)(\overline L R)$  and $(\overline L R)(\overline  R L)$ $\psi^4$ operators.}    and a factor of $g$ into $X$. In our actual calculations, we will revert to the unrescaled original operator basis $Q_i$.

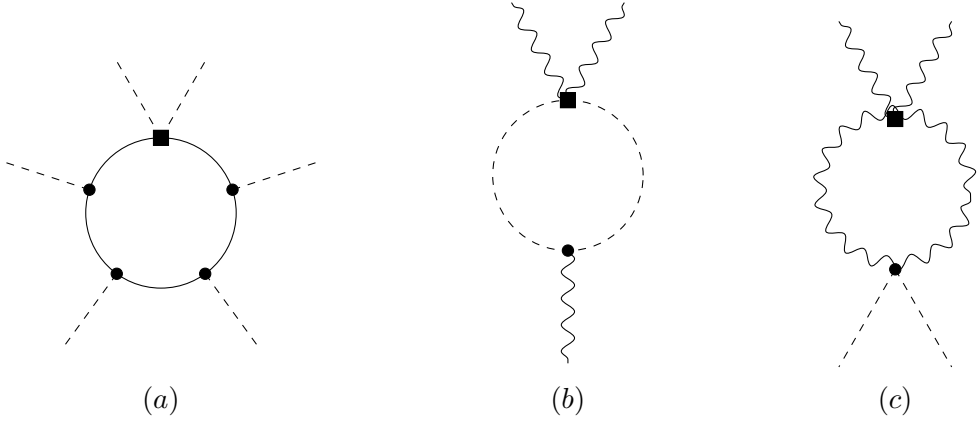
\begin{figure}
\centering
%
\begin{tikzpicture}

\draw (0,0) circle (1);

\filldraw (-0.1,0.9) rectangle (0.1,1.1);

\filldraw (162:1) circle (0.075);
\filldraw (234:1) circle (0.075);
\filldraw (306:1) circle (0.075);
\filldraw (18:1) circle (0.075);

\draw[dashed] (90:1) -- +(60:1.25) ;
\draw[dashed] (90:1) -- +(120:1.25) ;

\draw[dashed] (162:1) -- +(162:1.25) ;

\draw[dashed] (234:1) -- +(234:1.25) ;

\draw[dashed] (306:1) -- +(306:1.25);

\draw[dashed] (18:1) -- +(18:1.25);

\draw (0,-2.5) node [align=center] {$(a)$};

\end{tikzpicture}
\hspace{2cm}
%
\begin{tikzpicture}

\draw[dashed] (0,0) circle (1);

\filldraw (-0.1,0.9) rectangle (0.1,1.1);

\filldraw (270:1) circle (0.075);

\draw[decorate,decoration={snake}] (90:1) -- +(60:1.5) ;
\draw[decorate,decoration={snake}] (90:1) -- +(120:1.5) ;

\draw[decorate,decoration={snake}] (270:1) -- +(270:1.5);

\draw (0,-3.0) node [align=center] {$(b)$};

\end{tikzpicture}
\hspace{2cm}
\begin{tikzpicture}

\draw[decorate,decoration=snake]  (0,0) circle (1);

\filldraw (-0.1,0.9) rectangle (0.1,1.1);

\filldraw (270:1) circle (0.075);

\draw[decorate,decoration=snake]  (90:1) -- +(60:1.5) ;
\draw[decorate,decoration=snake]  (90:1) -- +(120:1.5) ;

\draw[dashed]  (270:1) -- +(240:1.5) ;
\draw[dashed]  (270:1) -- +(300:1.5) ;

\draw (0,-2.75) node [align=center] {$(c)$};

\end{tikzpicture}
\caption{\label{fig:1} (a) A $H^6 - \psi^2 H^2 D$ anomalous dimension graph which vanishes. (b) A $g^3 X^3-g^2X^2 H^2$ anomalous dimension graph of order $1$. (c) A $g^2X^2 H^2- g^3 X^3$ anomalous dimension graph of order $g^4$.
The solid square is a vertex from $\lsix$ and the dots are vertices from $\mathcal{L}_{\rm SM}$. }
\end{figure}

The anomalous dimension matrix (for the coefficients) in the rescaled basis $\widetilde Q_i$ has the form
shown in Table~\ref{tab:anom}, where we have given the explicit operator rescaling.
\begin{table}
\renewcommand{\arraycolsep}{0.15cm}
\renewcommand{\arraystretch}{1.5}
\begin{align*}
\begin{array}{cc|cccccccc}
&&g^3 X^3 & H^6 & H^4 D^2 & g^2 X^2 H^2 & y \psi^2 H^3 & g y \psi^2 X H & \psi^2 H^2 D & \psi^4 \\
&& 1 & 2 & 3 & 4 & 5 & 6 & 7 & 8\\
\hline
g^3 X^3 & 1 & 0 & 0 & 0 &1 & 0 & 0 & 0 & 0 \\
H^6 & 2 & g^6\lambda & 0 & g^2\lambda,\lambda^2 & \lambda g^4 & \lambda y^2 & 0 & \lambda g^2, \lambda y^2 & 0 \\
H^4 D^2 & 3 & g^6 & 0 & g^2  &g^4  & 0 & g^2y^2 & g^2 & 0 \\
g^2 X^2 H^2 & 4 &  g^4 & 0 & 0 &  0  & 0 & 0 & 0 & 0 \\
y \psi^2 H^3 & 5 & g^6 & 0 & g^2,\lambda,y^2 &  g^4  & y^2 &  g^2 \lambda, g^2y^2 & g^2,\lambda,y^2 & \lambda,y^2 \\
g  y \psi^2 X H & 6 & g^4  & 0 & 0 &  0 & 0 & g^2, y^2 & 1  &  1  \\
\psi^2 H^2 D & 7 & g^6 & 0 & g^2 &  g^4  & 0 & g^2y^2 & g^2,y^2 & g^2,y^2 \\
\psi^4 & 8 & g^6 & 0 & 0 &  0 & 0 & g^2y^2 & g^2,y^2 & g^2,y^2 \\
\end{array}
\end{align*}
\begin{align*}
\begin{array}{cc|cccccccc}
&&g^3 X^3 & H^6 & H^4 D^2 & g^2 X^2 H^2 & y \psi^2 H^3 & g y \psi^2 X H & \psi^2 H^2 D & \psi^4 \\
&& 1 & 2 & 3 & 4 & 5 & 6 & 7 & 8\\
\hline
g^3 X^3 & 1 & g^2 & 0 & 0 & 1 & 0 & 0 & 0 & 0 \\
H^6 & 2 & 0 & \lambda, g^2 & g^4,g^2 \lambda,\lambda^2 & g^6, g^4\lambda  & y^4 & 0 & y^4 & 0 \\
H^4 D^2 & 3 & 0 & 0 & g^2 ,\lambda &  g^4 & y^2 & 0 & y^2 & 0 \\
g^2 X^2 H^2 & 4 &  g^4 & 0 & 1 &  g^2,\lambda & 0 & y^2 & 1 & 0 \\
y \psi^2 H^3 & 5 & 0 & 0 & g^2,y^2 &  g^4  &g^2, \lambda,y^2 & g^2\lambda ,g^4,g^2y^2 & g^2 , \lambda  , y^2 & y^2 \\
g  y \psi^2 X H & 6 & g^4  & 0 & 0 &  g^2 & 1 & g^2,y^2 & 1 & 1 \\
\psi^2 H^2 D & 7 & 0 & 0 & y^2 &  g^4 & y^2 & g^2y^2 & g^2,\lambda,y^2 & y^2 \\
\psi^4 & 8 & 0 & 0 & 0 &  0 & 0 & g^2y^2 & y^2 & g^2,y^2 \\
\end{array}
\end{align*}
\caption{\label{tab:anom} The form of the one-loop anomalous dimension matrix for the coefficients of dimension six operators in the rescaled basis. The rows and columns are labelled by the eight operator classes. The lower table gives entries for which there is a direct contribution from a one-particle irreducible one-loop graph. The upper table gives entries which are generated indirectly by using EOM, and for which there need not be a direct contributing graph. There are also $y^2$ contributions to all diagonal entries except $11$ from wavefunction renormalization.  In some cases, the graphs vanish or produce an EOM operator that is shifted to other terms, and the entry is zero.}
\end{table}
The lower table shows the entries given by direct computation of graphs, and the upper table shows entries that are possible by computing a graph and then converting it to the standard basis using EOM. The zero entries are those for which there is no one-loop divergent graph. The possible orders in $g$, $\lambda$ and $y$ are shown for the other entries. In some cases, the allowed graphs have vanishing divergence, such as Fig~\ref{fig:1}(a), so not all of  the possible terms in the array are non-zero. Formally including factors of $y$ into the the $\psi^2 H^3$ and $\psi^2 X H$ terms makes the matrix a function only of even powers of $y$.
The anomalous dimension matrix for coefficients of rescaled operators contains terms of order unity. For example, the graph in Fig.~\ref{fig:1}(b) gives an order $1$ contribution to the $g^3 X^3-g^2X^2 H^2$ entry. In terms of the original operators, the graph has one gauge coupling at the Standard Model vertex, and is order $g$. On rescaling, the $X^2 H^2$ operator at the $\lsix$ vertex gets a  factor of $g^2$, and the $X^3$ operator given by the external lines absorbs a factor of $g^3$, so the graph becomes order $g \times g^2/g^3 = 1$. Similarly graph Fig.~\ref{fig:1}(c), which is order $g^3$ in terms of the original operators ($g^2$ from the standard model vertex and $g$ from the $\lsix$ vertex), is order $g^4$ in terms of the rescaled operators.

All the entries in the one-loop anomalous dimension matrix contain the usual $1/(16\pi^2)$ factor of a perturbative one-loop graph. However, there  are entries of order $1$, $g^2$, $y^2$, $g^4$, $y^4$, etc.\ so that it appears that the anomalous dimension matrix does not have the usual form, a product of powers of $g^2/(16\pi^2)$, $\lambda/(16\pi^2)$ and $y^2/(16\pi^2)$, with no extraneous factors of $16\pi^2$. To understand what is going on, it is instructive to consider the rescaled operators normalized using  naive dimensional analysis~\cite{Manohar:1983md}. The general Lagrangian term is
\begin{align}
f^2 \Lambda^2 \left(\frac{\psi}{f \sqrt \Lambda}\right)^a
\left(\frac{H}{f}\right)^b\left(\frac{y H}{\Lambda}\right)^c \left(\frac{D}{\Lambda}\right)^d \left( \frac{g X}{\Lambda^2} \right)^e
\label{5.2}
\end{align}
with $\Lambda \sim 4 \pi f$. The $H$ and $y H$ terms have the same scaling if $y \sim 4\pi$. If $y < 4\pi $, then one gets the usual suppression of chirality flip terms in weak coupling, analogous to the suppression of gauge interactions in weak coupling discussed in Ref.~\cite{Manohar:1983md}. The eight operator classes give
\begin{align}
\frac{f^2}{\Lambda^4}\ g^3 X^3,\ \frac{\Lambda^2}{f^4}\ H^6,\ \frac{1}{f^2}\ H^4 D^2,\ \frac{1}{\Lambda^2}\ g^2 X^2 H^2,\
\frac{1}{f^2}\ y\psi^2 H^3,\ \frac{1}{\Lambda^2}\ y\psi^2 g X H,\ \frac{1}{f^2} \psi^2 H^2 D,\ \frac{1}{f^2} \psi^4
\label{3.5}
\end{align}
times coefficients of order one for $\lsix$.

Let $\widehat Q_i$ be the $\lsix$ operators normalized as in Eq.~(\ref{3.5}), so that their coefficients $\widehat C_i$ are dimensionless, and expected to be order unity by naive dimensional analysis. Then one sees that the contribution of graph Fig.~\ref{fig:1}(b,c), can be written in three equivalent ways,
\begin{align}
\mu \frac{\rd}{\rd \mu}  C_1 &=  \frac{A_b}{16\pi^2}\, g\,  C_4, & \mu \frac{\rd}{\rd \mu}  C_4 &=  \frac{A_c}{16\pi^2}\, g^3\,  C_1, \nn
\mu \frac{\rd}{\rd \mu} \widetilde C_1 &=  \frac{A_b}{16\pi^2}\, \widetilde C_4, & \mu \frac{\rd}{\rd \mu} \widetilde C_4 &=  \frac{A_c}{16\pi^2}\,
g^4\, \widetilde C_1, \nn
\mu \frac{\rd}{\rd \mu} \widehat C_1 &=  A_b\, \widehat C_4, & \mu \frac{\rd}{\rd \mu} \widehat C_4 &=   A_c \left(\frac{g^2}{16\pi^2}\right)^2 \, \widehat C_1\,,
\end{align}
where $A_{b,c}$ are constants.
One can see from the last row that, with the normalization Eq.~(\ref{3.5}), the anomalous dimension for $\widehat C_i$ depends on products of powers of
$ \lambda/(16\pi^2)^2$, $g^2/(16\pi^2)$, and $y^2/(16\pi^2)$ as expected. It is straightforward to verify this for the  entire matrix. Terms such as $\gamma_{14}$ are order unity and effectively zeroth order, $\gamma_{44}$ is of one-loop size, $\gamma_{41}$ of two-loop size, $\gamma_{31}$ of three-loop size, and $\gamma_{21}$ of four-loop size, etc., even though all of them are given by one-loop diagrams in the EFT.

It is worth emphasizing that, while the the use of Eq.~(\ref{3.5}) for the normalization makes it easier to understand the importance of various terms, it does not affect the actual calculation. One can convert from one normalization to another using the trivial rescaling in Eq.~(\ref{3.3}).
 When we refer to anomalous dimension entries as order $g^2$, etc.\ we will use the rescaled form in Table~\ref{tab:anom} in either the $\hat C_i$ or $\widetilde C_i$ normalization, which differ only by factors of $4\pi$.
 The explicit RGE are given for the original unrescaled coefficients $C_i$.

The effects of $\lsix$ are suppressed by $1/\Lambda^2$, and vanish as $\Lambda \to \infty$, so the RGE does not need to be integrated over a large range of $t=\ln \mu$. The integration can be done in perturbation theory by expanding in powers of the anomalous dimension matrix $\gamma$.  Dropping $\beta$-function running of the couplings for simplicity,
\begin{align}
C(t) &= \left[1+ t \gamma +\frac12 t^2 \gamma \cdot \gamma + \ldots \right] C(0)\,.
\end{align}
Different powers of $\gamma$ can contribute at the same order, because of the  structure of Table~\ref{tab:anom}. For example, a one-loop contribution of order $\lambda/(16\pi^2)$ is generated by the product of the order $\lambda$ term in $\gamma_{33}$ and the order 1 term in $\gamma_{43}$ at second order in $\gamma$.
To get the coefficients of all 59 operators accurate to one-loop order (i.e.\ including all $g^2/(16\pi^2)$, $\lambda/(16\pi^2)$ and $y^2/(16\pi^2)$ corrections) requires keeping terms to third order in $\gamma$.

The operator $X^2 H^2$ contributes to $h \to \gamma \gamma$ and $h \to \gamma Z$, which are one-loop amplitudes in the Standard Model. In Ref.~\cite{Grojean:2013kd}, we restricted our attention to the $X^2 H^2$ operators, and computed the $8 \times 8$ $\gamma_{44}$ submatrix of the $59 \times 59$ anomalous dimension matrix. The largest effects were due to the $\lambda H^4$ and Yukawa couplings, rather than the gauge couplings.

In this paper, we give the results for terms in the one-loop anomalous dimension matrix that depend only on $\lambda$, and are independent of the gauge couplings, i.e.\ the terms of order $\lambda$, $\lambda^2$, and $\lambda y^2$ in Table~\ref{tab:anom}.
The remaining terms will be discussed in subsequent publications. Note that the results in Sec.~\ref{sec:smrge} keep the full $g,\lambda,y$ dependence at one-loop, and do not drop any terms.

\begin{figure}
\centering
\begin{tikzpicture}

\draw (0,0) circle (1);

\filldraw (-0.1,0.9) rectangle (0.1,1.1);

\filldraw (210:1) circle (0.075);
\filldraw (330:1) circle (0.075);

\draw (90:1) -- +(60:1.5) ;
\draw (90:1) -- +(120:1.5) ;

\draw[dashed]  (210:1) -- +(210:1.5);

\draw[decorate,decoration=snake] (330:1) -- +(330:1.5);

\end{tikzpicture}
\caption{\label{fig:68} Diagram contributing to the $\psi^2 X H- \psi^4$ anomalous dimension $\gamma_{68}$ given in Eq.~(\ref{68}).
The solid square is a $\psi^4$ vertex from $\lsix$ and the dots are gauge and Yukawa vertices from $\mathcal{L}_{\rm SM}$. }
\end{figure}
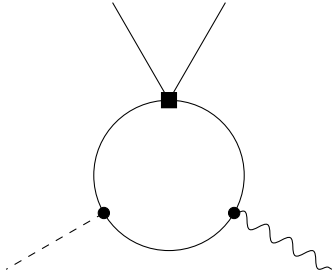

There are terms in $\gamma$ of order 1. These arise from graphs such as Fig.~\ref{fig:68} involving gauge fields.
The graph is order $g y$ because it has one gauge, and one Yukawa vertex, but becomes order 1 in our rescaled basis. We will discuss these terms in a subsequent paper. Here we give an example of one such term, $\gamma_{68}$, the mixing of four-fermion operators
with magnetic moment operators,
\begin{align*}
\mu \frac{\rd}{\rd \mu} C_{\substack{eB \\ pr}} &=  \frac{1}{16\pi^2}\left[ 4g_1N_c\, (\hyp_u+\hyp_q) C_{\substack{lequ \\ prst}}^{(3)}\, [Y_u]_{ts} \right]+\ldots \, \nn
\mu \frac{\rd}{\rd \mu} C_{\substack{eW \\ pr}} &=  \frac{1}{16\pi^2}\left[- 2g_2 N_c\, C_{\substack{lequ \\ prst}}^{(3)}\, [Y_u]_{ts} \right]+ \ldots \,
\end{align*}
\vspace{-0.5cm}
\begin{align}
\mu \frac{\rd}{\rd \mu} C_{\substack{uB \\ pr}} &=  \frac{1}{16\pi^2}\left[4g_1(\hyp_e+\hyp_l)\, C_{\substack{lequ \\ stpr}}^{(3)}\, [Y_e]_{ts} \right]+ \ldots \, \nn
\mu \frac{\rd}{\rd \mu} C_{\substack{uW \\ pr}} &= \frac{1}{16\pi^2}\left[-2g_2 C_{\substack{lequ \\ stpr}}^{(3)}\,  [Y_e]_{ts} \right]+ \ldots\,,
\label{4.3}
\end{align}
where $\ldots$ denotes  contributions from other operators, and $\hyp_i$ are the $U(1)$ hypercharges.

Eq.~(\ref{4.3}) is an example of non-zero mixing between ``tree'' and ``loop'' operators.
Eq.~(\ref{4.3}) cannot be cancelled by other terms, since there are no redundant operators in the basis we use. The operator $Q_{lequ}^{(3)}$ can be Fierzed into scalar form ($\alpha$ is a color index),
\begin{align}
Q_{lequ}^{(3)} &= (\bar l_p^j \sigma_{\mu\nu} e_r) \epsilon_{jk} (\bar q_s^k \sigma^{\mu\nu} u_t) =-4  (\bar l_p^j e_r) \epsilon_{jk} (\bar q_s^{k \alpha}  u_{\alpha t} ) - 8  (\bar l_p^j u_{\alpha t}) \epsilon_{jk} (\bar q_s^{k\alpha}  e_r) \nn
&=-4 Q_{lequ}^{(1)}  - 8  (\bar l_p^j u_{\alpha t}) \epsilon_{jk} (\bar q_s^{k\alpha}  e_r)
\label{68}
\end{align}
and can be generated by the tree-level exchange of $({\bf 3},{\bf 2},7/6)$ scalars, i.e.\ those with the quantum numbers of a leptoquark doublet. Tree-level exchange of leptoquarks and heavy $({\bf 1},{\bf 2},1/2)$ scalars with $H$-field quantum numbers can generate any combination of $Q_{lequ}^{(1)}$ and
$Q_{lequ}^{(3)}$.

\section{$\lambda, \lambda^2, \lambda y^2$ Contributions to the $\lsix$ Anomalous Dimension Matrix}\label{sec:lam}

\begin{figure}
\centering
\begin{tikzpicture}

\draw[dashed] (0,0) circle (1);

\filldraw (-0.1,0.9) rectangle (0.1,1.1);

\filldraw (270:1) circle (0.075);

\draw[dashed]  (90:1) -- +(60:1.5) ;
\draw[dashed]  (90:1) -- +(120:1.5) ;

\draw[dashed]  (270:1) -- +(240:1.5) ;
\draw[dashed]  (270:1) -- +(300:1.5) ;

\end{tikzpicture}
\caption{\label{fig:2} Graph contributing to the $H^4 D^2-H^4D^2$ anomalous dimension and to EOM operators.
The solid square is a $H^4 D^2$ vertex from $\lsix$ and the dot is the $\lambda (H^\dagger H)^2$  vertex from $\mathcal{L}_{\rm SM}$. }
\end{figure}
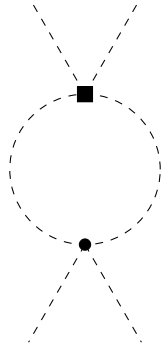

The computation of the $\lambda, \lambda^2, \lambda y^2$ anomalous dimensions has some subtleties. An example is the graph in Fig.~\ref{fig:2}
which generates, in addition to the $Q_{H\Box}$ and $Q_{H D}$ operators, the EOM operator $E_{H\Box}$ of Eq.~(\ref{ebox}). Eq.~(\ref{ebox2}) eliminates $E_{H \Box}$ in terms of our standard basis of operators, so Fig.~\ref{fig:2} contributes to the running of the $H^6$ coefficient $C_H$, as well as the $\psi^2 H^3$ coefficients $C_{uH}$, $C_{dH}$ and $C_{eH}$, and to the running of the dimension four SM coefficients in Eq.~(\ref{smrge}). Fig.~\ref{fig:2} is an example of how terms get shuffled around by the EOM. Fig.~\ref{fig:2} has only external $H$ fields, but contributes to the running of the $\psi^2 H^3$ operators.

The equations presented below are \emph{not the complete} RGE, but only the $\lambda, \lambda^2, \lambda y^2$ terms. The remaining terms are lengthy, and will be given a subsequent publication.
The evolution of the $H^6$ coefficient is
\begin{align}
\mu \frac{\rd}{\rd \mu} C_H &= \frac{1}{16\pi^2} \left[ 108\, \lambda\, C_H -160\, \lambda^2\, C_{H \Box}+48\, \lambda^2\, C_{H D} \right]+\frac{8\lambda}{16\pi^2}  \eta_1 + \frac{8\lambda}{16\pi^2}  \eta_2
\end{align}
where $\eta_{1,2}$ are given in Eq.~(\ref{etadef}). The diagonal $C_H-C_H$ term
$108\lambda/(16\pi^2)$ has a large numerical coefficient, and is independent of the normalization chosen for the $H^6$ operator, i.e.\ whether we use $(H^\dagger H)^3$ or $(H^\dagger H)^3/(3!)^2$, etc.\ The large number 108 arises from the combinatorics of the Wick contractions. For $m_H \sim 126$~GeV,
$108\,\lambda/(16\pi^2) \approx 0.1$.

The evolution of the $X^2 H^2$ coefficients  is
\begin{align}
\mu \frac{\rd}{\rd \mu} C_{H G} &= \frac{12\lambda}{16\pi^2}\, C_{HG}\,, &
\mu \frac{\rd}{\rd \mu} C_{H \widetilde G} &= \frac{12\lambda}{16\pi^2}\,  C_{H \widetilde G} \,,\nn
\mu \frac{\rd}{\rd \mu} C_{H W} &= \frac{12 \lambda}{16\pi^2}\, C_{HW}\,, &
\mu \frac{\rd}{\rd \mu} C_{H \widetilde W} &= \frac{12\lambda}{16\pi^2}\,  C_{H \widetilde W}\,, \nn
\mu \frac{\rd}{\rd \mu} C_{H B} &= \frac{12\lambda}{16\pi^2}\,  C_{HB}\,, &
\mu \frac{\rd}{\rd \mu} C_{H \widetilde B} &= \frac{12 \lambda}{16\pi^2}\,   C_{H \widetilde B} \,, \nn
\mu \frac{\rd}{\rd \mu} C_{H WB} &= \frac{4 \lambda}{16\pi^2}\,  C_{HWB}\,, &
\mu \frac{\rd}{\rd \mu} C_{H \widetilde WB} &= \frac{4 \lambda}{16\pi^2}\, C_{H \widetilde WB}\,,
\end{align}
and is part of the complete $\gamma_{44}$ calculation given previously in Ref.~\cite{Grojean:2013kd}.

The $H^4 D^2$ terms are
\begin{align}
\mu \frac{\rd}{\rd \mu} C_{H\Box} &= \frac{24\lambda}{16\pi^2} \, C_{H \Box}\,,  &
\mu \frac{\rd}{\rd \mu} C_{HD} &= \frac{12\lambda}{16\pi^2} \, C_{H D}  \,,
\end{align}
and the $\psi^2 H^3$ terms are
\begin{align}
\mu \frac{\rd}{\rd \mu} C_{\substack{uH \\ rs}}  &= \frac{\lambda}{16\pi^2}  \biggl[ 24\, C_{\substack{uH  \\ rs}}-4 C^{(1)}_{\substack{Hq \\ rt}}\, [Y_u^\dagger]_{ts}
+12 C^{(3)}_{\substack{Hq \\ rt}}\, [Y_u^\dagger]_{ts} + 4  [Y_u^\dagger]_{rt}\, C_{\substack{Hu \\ ts}}  - 4  [Y_d^\dagger]_{rt}\, C^*_{\substack{Hud \\ st}} \nn
& -4\, [Y_u^\dagger]_{rs} C_{H \Box} +2 [Y_u^\dagger]_{rs} C_{HD}
-8 C^{(1)}_{\substack{qu \\ rpts}} [Y_u^\dagger]_{pt}-8 c_{F,3} C^{(8)}_{\substack{qu \\ rpts}} [Y_u^\dagger]_{pt}
 -4 C^{(1)}_{\substack{lequ \\ ptrs}} [Y_e]_{tp} \nn
 &+4 N_c C^{(1)}_{\substack{quqd \\ rspt}} [Y_d]_{tp}+2 C^{(1)}_{\substack{quqd \\ psrt}} [Y_d]_{tp}
 +2 c_{F,3} C^{(8)}_{\substack{quqd \\ psrt}} [Y_d]_{tp}\biggr]\,, \nn
\mu \frac{\rd}{\rd \mu} C_{\substack{dH \\ rs}}  &= \frac{\lambda}{16\pi^2} \biggl[24\, C_{\substack{dH  \\ rs}}+4 C^{(1)}_{\substack{Hq \\ rt}}\, [Y_d^\dagger]_{ts}
+12 C^{(3)}_{\substack{Hq \\ rt}}\, [Y_d^\dagger]_{ts} - 4  [Y_d^\dagger]_{rt}\, C_{\substack{Hd \\ ts}} - 4  [Y_u^\dagger]_{rt}\, C_{\substack{Hud \\ ts}}  \nn
& -4\,  [Y_d^\dagger]_{rs}  C_{H \Box} +2  [Y_d^\dagger]_{rs}  C_{HD}
-8 C^{(1)}_{\substack{qd \\ rpts}} [Y_d^\dagger]_{pt} - 8 c_{F,3} C^{(8)}_{\substack{qd \\ rpts}} [Y_d^\dagger]_{pt} + 4  C^*_{\substack{ledq \\ ptsr}} [Y_e^\dagger]_{pt} \nn
& +4 N_c C^{(1)}_{\substack{quqd \\ ptrs}} [Y_u]_{tp}+ 2 C^{(1)}_{\substack{quqd \\ rtps}} [Y_u]_{tp}+ 2 c_{F,3} C^{(8)}_{\substack{quqd \\ rtps}} [Y_u]_{tp}
\biggr]\,,  \displaybreak \nn
\mu \frac{\rd}{\rd \mu} C_{\substack{eH \\ rs}}  &= \frac{\lambda}{16\pi^2}  \biggl[ 24\, C_{\substack{eH  \\ rs}}+4 C^{(1)}_{\substack{Hl \\ rt}}\, [Y_e^\dagger]_{ts}
+12 C^{(3)}_{\substack{Hl \\ rt}}\, [Y_e^\dagger]_{ts} - 4  [Y_e^\dagger]_{rt}\, C_{\substack{He \\ ts}}\nn
&-4\, [Y_e^\dagger]_{rs} C_{H \Box} +2 [Y_e^\dagger]_{rs} C_{HD}
-8 C_{\substack{le \\ rpts}} [Y_e^\dagger]_{pt} + 4 N_c C_{\substack{ledq \\ rspt}} [Y_d^\dagger]_{tp} -4 N_c  C^{(1)}_{\substack{lequ \\ rspt}} [Y_u]_{tp} \biggr]\,,
\end{align}
There are no other one-loop $\lambda$, $\lambda^2$ and $\lambda y^2$ terms.

\section{Conclusions}

We have given the structure of the $59 \times 59$ anomalous dimension matrix for dimension-six operators in the Standard Model, and presented all the terms of order $\lambda$, $\lambda^2$ and $\lambda y^2$ that can arise at one loop. We have also given one example of tree-loop mixing among the dimension-six operators. The remaining one-loop terms will be discussed in a subsequent publication. In addition, we have  given the full contribution of $\lsix$ to the RGE of the usual dimension-four terms and the dimension-two term $H^\dagger H$ in the Standard Model Lagrangian.

{\it Note added:} While this paper was being readied for publication, Ref.~\cite{Elias-Miro:2013mua} appeared, which also discusses the anomalous dimension of $\lsix$. Ref.~\cite{Elias-Miro:2013mua} gives the full $\lambda,y,g$ dependence of a subset of the anomalous dimension matrix.
A different operator basis including 5 redundant operators is used, as well as a ``tree-loop'' analysis, so it is difficult to make a quick comparison of the common terms between the two calculations, but an initial look shows good agreement.
This work was supported in part by DOE grant DE-SC0009919. MT thanks the hospitality of the KITP, which is supported in part by the National Science Foundation under Grant No. PHY11-25915. \bibliographystyle{JHEP}
\bibliography{RG}

\end{document}